\documentclass[12pt]{article}
\usepackage{epsfig}

\newcommand\pubnumber{%RADCOR-2000-001
}
\newcommand\pubdate{\today}
\newcommand\hepnumber{hep-ph/0101152}

\def\csumb{Nuclear Physics Institute of Moscow State University\\
Moscow 119899, Russia}
\def\support{\footnote{
The participation in the Symposium supported by the Organizing
Committee and by the Russian State Research Program
`High Energy Physics'. Work supported by the Volkswagen Foundation,
contract No.~I/73611, and by the Russian Foundation for Basic
Research, project 98--02--16981.
}}

\def\Title#1{\begin{center} {\Large\bf #1 } \end{center}}
\def\Author#1{\begin{center}{ \sc #1} \end{center}}
\def\Address#1{\begin{center}{ \it #1} \end{center}}

\newcommand\pubblock{\rightline{\begin{tabular}{l} \pubnumber\\
         \pubdate\\ \hepnumber \end{tabular}}}
\newenvironment{Abstract}{\begin{quotation}  }{\end{quotation}}
\newenvironment{Presented}{\begin{quotation} \begin{center}
             Presented at the\end{center}
      \begin{center}\begin{large}}{\end{large}\end{center} \end{quotation}}
\def\Acknowledgments{\bigskip  \bigskip \begin{center}
          \large\bf Acknowledgments\end{center}}

\makeatletter
\def\section{\@startsection{section}{0}{\z@}{5.5ex plus .5ex minus
 1.5ex}{2.3ex plus .2ex}{\large\bf}}
\def\subsection{\@startsection{subsection}{1}{\z@}{3.5ex plus .5ex minus
 1.5ex}{1.3ex plus .2ex}{\normalsize\bf}}
\def\subsubsection{\@startsection{subsubsection}{2}{\z@}{-3.5ex plus
-1ex minus  -.2ex}{2.3ex plus .2ex}{\normalsize\sl}}

\renewcommand{\@makecaption}[2]{%
   \vskip 10pt
   \setbox\@tempboxa\hbox{\small #1: #2}
   \ifdim \wd\@tempboxa >\hsize     % IF longer than one line:
       \small #1: #2\par          %   THEN set as ordinary paragraph.
     \else                        %   ELSE  center.
       \hbox to\hsize{\hfil\box\@tempboxa\hfil}
   \fi}

 \def\citenum#1{{\def\@cite##1##2{##1}\cite{#1}}}
\newcount\@tempcntc
\def\@citex[#1]#2{\if@filesw\immediate\write\@auxout{\string\citation{#2}}\fi
  \@tempcnta\z@\@tempcntb\m@ne\def\@citea{}\@cite{\@for\@citeb:=#2\do
    {\@ifundefined
       {b@\@citeb}{\@citeo\@tempcntb\m@ne\@citea\def\@citea{,}{\bf ?}\@warning
       {Citation `\@citeb' on page \thepage \space undefined}}%
    {\setbox\z@\hbox{\global\@tempcntc0\csname b@\@citeb\endcsname\relax}%
     \ifnum\@tempcntc=\z@ \@citeo\@tempcntb\m@ne
       \@citea\def\@citea{,}\hbox{\csname b@\@citeb\endcsname}%
     \else
      \advance\@tempcntb\@ne
      \ifnum\@tempcntb=\@tempcntc
      \else\advance\@tempcntb\m@ne\@citeo
      \@tempcnta\@tempcntc\@tempcntb\@tempcntc\fi\fi}}\@citeo}{#1}}
\def\@citeo{\ifnum\@tempcnta>\@tempcntb\else\@citea\def\@citea{,}%
  \ifnum\@tempcnta=\@tempcntb\the\@tempcnta\else
  {\advance\@tempcnta\@ne\ifnum\@tempcnta=\@tempcntb \else\def\@citea{--}\fi
    \advance\@tempcnta\m@ne\the\@tempcnta\@citea\the\@tempcntb}\fi\fi}
\makeatother

\def\beq{\begin{equation}}
\def\eeq#1{\label{#1}\end{equation}}
\def\eeqn{\end{equation}}

\newenvironment{Eqnarray}%
   {\arraycolsep 0.14em\begin{eqnarray}}{\end{eqnarray}}
\def\beqa{\begin{Eqnarray}}
\def\eeqa#1{\label{#1}\end{Eqnarray}}
\def\eeqan{\end{Eqnarray}}

\let\bar=\overbar

\def\Dslash{\not{\hbox{\kern-4pt $D$}}}
\def\dslash{\not{\hbox{\kern-2pt $\del$}}}

\def\msb{{\bar{\ssstyle M \kern -1pt S}}}

\def\lsim{\mathrel{\raise.3ex\hbox{$<$\kern-.75em\lower1ex\hbox{$\sim$}}}}
\def\gsim{\mathrel{\raise.3ex\hbox{$>$\kern-.75em\lower1ex\hbox{$\sim$}}}}

\newcommand{\be}{\begin{equation}}
\newcommand{\ee}{\end{equation}}
\newcommand{\bea}{\begin{eqnarray}}
\newcommand{\eea}{\end{eqnarray}}
\newcommand{\beas}{\begin{subeqnarray}}
\newcommand{\eeas}{\end{subeqnarray}}
\newcommand{\al}{\alpha}

\newcommand{\gm}{\gamma}
\newcommand{\Gm}{\Gamma}

\newcommand{\eps}{\varepsilon}

\newcommand{\ep}{\varepsilon}

\newcommand{\om}{\omega}

\newcommand{\dd}{\mbox{d}}

\newcommand{\lra}{\leftrightarrow}

\newcommand{\uk}{\underline{k}}

\newcommand{\I}{i}
\newcommand{\E}{e}
\newcommand{\nn}{\nonumber}

\newcommand{\cdo}{\!\cdot\!}

\newcommand{\ria}{\rightarrow}

\newcommand{\Li}[2]{{\mbox{Li}}_{#1}\left(#2\right)}

\begin{document}
\begin{titlepage}
\pubblock

\vfill
\def\thefootnote{\fnsymbol{footnote}}
\Title{`Strategy of Regions': \\
 Expansions of Feynman Diagrams  \\%[5pt]
 both in Euclidean and \\
 Pseudo-Euclidean Regimes}
\vfill
\Author{V.A. Smirnov\support}
\Address{\csumb}
\vfill
\begin{Abstract}
The strategy of regions [1] turns out to be a universal method for
expanding  Feynman
integrals in various limits of momenta and masses. This strategy
is reviewed and
illustrated through numerous examples.
In the case of typically Euclidean
limits it is equivalent to well-known prescriptions within the strategy
of subgraphs.
For regimes typical for Minkowski space, where the strategy of subgraphs
has not yet been developed, the strategy of regions is characterized
in the case of threshold limit, Sudakov limit and Regge limit.
\end{Abstract}
\vfill
\begin{Presented}
5th International Symposium on Radiative Corrections \\
(RADCOR--2000) \\[4pt]
Carmel CA, USA, 11--15 September, 2000
\end{Presented}
\vfill
\end{titlepage}
\def\thefootnote{\arabic{footnote}}
\setcounter{footnote}{0}

\section{Introduction}

The problem of asymptotic\footnote{The word `asymptotic' is also usually
applied to perturbative series with zero radius of
convergence.
For expansions of Feynman integrals in momenta and masses,
this word just means that the remainder
of an asymptotic expansion satisfies a desired estimate provided we
pick up a sufficiently large number of first terms of the expansion.
It should be stressed that the radius of convergence of any series in
the right-hand side of any expansion in momenta and masses is
non-zero. This is not a rigorously proven mathematical theorem but at
least examples where such a radius of convergence is zero are unknown for
the moment.}
expansions of Feynman integrals in momenta
masses is very important and has been analyzed in a large number of papers.
For limits typical for Euclidean space, an adequate solution has been
found \cite{SS} (see a brief review in \cite{Sm3}) and
mathematically proven.
It is expressed by a simple formula with summation in a certain family
of subgraphs of a given graph so that let us refer to it as
`the strategy of subgraphs'.
For limits typical for Minkowski space, the strategy of subgraphs
has not yet been rigorously developed.

Quite recently a new  method for expanding  Feynman
integrals in limits of momenta and masses has been suggested \cite{BS}.
It is based on the analysis of various regions in the space of
loop momenta of a given diagram and denoted as `the strategy of
regions'. The purpose of this talk is to
review and
illustrate this strategy through numerous examples.
First, the problem of asymptotic expansion in limits of momenta and
masses is characterized. Then the two basic strategies are formulated
and compared for limits typical for
Euclidean space.
For regimes typical for Minkowski space, the strategy of regions
is checked
through typical examples, up to two-loop level, in the case of
threshold limit, Sudakov limit and Regge limit.
Finally, the present status of the strategy of regions is
characterized.

\section{Limits of momenta and masses}

Let $\Gm$ be a graph and $F_{\Gm}(m_1,m_2,\ldots,q_1,q_2,\ldots)$
the corresponding Feynman integral constructed according to
Feynman rules and depending on masses $m_i$ and external momenta $q_j$.
It can be represented as a linear combination
of tensors composed of the external momenta with coefficients which
are scalar Feynman integrals that depend on the masses and
kinematical invariants $s_{ij} = q_i \cdo q_j$.

The problem of asymptotic expansion of Feynman integrals in some limit of
momenta and masses is of the physical origin and arises quite naturally.
If one deals with phenomena that take place at a given energy scale it is
natural to consider large (small) all the masses and kinematical invariants
that are above (below) this scale. Therefore a limit (regime) is nothing
but a decomposition of the given family of these parameters into small
and large ones.

For limits typical for Euclidean space, an external momentum is called
large if at least one of its components is large and small if all its
four components are large. Thus such a limit is characterized by
a decomposition
$\{m_i\},\{q_i \} \to \{m_i\},\{q_i \}; \{M_i\},\{Q_i \} $,
with $m_i, |q_j| \ll M_{i'}, |Q_{j'}|$,
where $|q_j|$ is understood in the Euclidean  sense.

For limits typical for pseudo-Euclidean space, it is impossible to
characterize the external momenta in this way and one turns to a decomposition
written through kinematical invariants:
$\{m_i\},\{s_{j j'} \} \to \{m_i\},\{s_{j j'} \};
\{M_i\},\{S_{j j'}\}$,
with $m_i, |s_{j j'}| \ll M_{i'}, |S_{j j'}|$. However, instead of
the kinematical invariants themselves, some linear combinations can be used
(for example, in the case of the threshold limit).

Feynman integrals are generally quite complicated functions depending
on a large number of arguments.
When a given Feynman integral is considered in a given limit it looks
natural to expand it in ratios of small and large parameters and then replace
the initial complicated object by a sufficiently large number of
first terms of the corresponding asymptotic
expansion.
Experience shows that Feynman integrals are always expanded in powers
and logarithms of the expansion parameter which is a ratio of the large and
the small scales of the problem. In particular, when a given Feynman integral
depends only on a small mass squared and a large external momentum squared,
$m^2 \ll -q^2$, we have
\be
F_{\Gm} (q^2,m^2)
\sim (-q^2)^{\om} \sum_{n=n_0}^{\infty} \sum_{j=0}^{2h}
C_{nj}\, \left(\frac{m^2}{-q^2} \right)^n
 \ln^j \left(\frac{m^2}{-q^2}\right)\; ,
\ee
where $h$ is the number of loops of $\Gm$ and  $\om$
ultraviolet (UV) degree of divergence. The maximal power of the logarithm equals
the number of loops for typically Euclidean limits and is twice the number
of loops for limits typical for Minkowski space.

%********************
To expand Feynman diagrams one can either
\begin{enumerate}
\item
Take a given diagram in a given limit and expand it by some special
technique, or,
\item
Formulate prescriptions for a given limit and then
apply them to any
diagram (e.g. with 100 loops).
\end{enumerate}

Of course, the second (global) solution is preferable because
\begin{itemize}
\item
{\em no analytical work} is needed when applying it to a given diagram:
just follow
formulated prescriptions and write down a result in terms of Feynman
integrals (with integrands expanded in Taylor series in some
parameters);
%\item
%individual terms in the expansion are evaluated without evaluating the full
%result;
\item
a natural requirement can be satisfied: individual terms of the expansion are
homogeneous (modulo logs) in the expansion parameter.
%;\item
%factorization of scales.
\end{itemize}

Two kinds of such global prescriptions are known:
\begin{itemize}
\item[*]
{\em Strategy of Subgraphs} and
\item[*]
{\em Strategy of Regions}
\end{itemize}

We shall now formulate both strategies
in the case of limits typical for Euclidean space.

\section{Strategy of subgraphs and strategy of regions for limits
typical for Euclidean space}

For limits typical for Euclidean space,
the solution of the problem of asymptotic expansion is described \cite{SS}
by the following simple formula, with summation in subgraphs, supplied
with some explanations:
\be
F_{\Gm} \sim \sum_{\gm}
F_{\Gm/\gm} \circ {\cal T}_{\gm}F_{\gm}\;,
\label{AE}
\ee
where
the sum runs in a certain class of subgraphs $\gm$ of
$\Gm$. For example, in the off-shell (Euclidean) limit $m^2 \ll -q^2$
(when $q$ is considered large in Euclidean sense),
one can
distribute the flow of $q$ through {\em all} the lines of
$\gm$. (This is a `physical' definition.)
Moreover
$F_{\gm}$ and $F_{\Gm/\gm}$ are the Feynman integrals
respectively  for $\gm$ and $\Gm/\gm$ (the reduced graph $\Gm/\gm$
is obtained from
$\Gm$ by collapsing $\gm$ to a point). The operator
${\cal T}_{\gm}$ expands the integrand of $F_{\gm}$ in Taylor
series in
its small masses and small external momenta which are either
the small external momenta of $\Gm$, or loop momenta of the
whole graph that are external for $\gm$ (they are {\em by definition}
small). The symbol
$\circ$ denotes insertion of the second factor (polynomial) into
$F_{\Gm/\gm}$ (like an insertion of a counterterm within dimensional
renormalization).

All quantities are supposed to be dimensionally  regularized
\cite{dimreg} by $d=4-2\ep$.
Even if the initial Feynman integral is UV and IR finite,
the regularization is necessary because individual terms in the right-hand
side become divergent starting from some minimal order of expansion.
The necessity to run into divergences is a negligible price
to have the simplest prescription for expanding Feynman integrals.
Moreover the cancellation of divergences in the right-hand
side of expansions of finite Feynman integrals is a very crucial
practical check of the expansion procedure.

Operator analogs of limits typical for Euclidean space
(the off-shell large momentum limit and the large mass limit)
are operator product expansion and large mass expansion
described by an effective Lagrangian --- see a review with applications
in \cite{CKK}.

\begin{figure}[b!]
\begin{center}
\epsfig{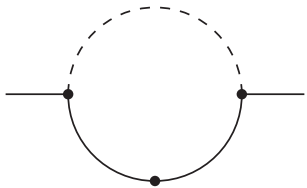}
\caption[0]{\label{} One-loop propagator diagram}
\label{fig1}
\end{center}
\end{figure}

Consider, for example, the scalar diagram shown in Fig.~\ref{fig1}
in the off-shell limit $m^2 \ll -q^2$ which can treated as a Euclidean
limit with the external momentum $q$ large in the Euclidean sense.
%%%%%%%%%%%%%%%%%%%%%%%%%%%%%%%%%%%%%%%%%%%%%%%%%%%%%%%%%%%%%%%%%%%%%%%%%%%
The propagator of the dashed line is massless and the dot
on the solid line denotes the second power of the massive
propagator.
The corresponding Feynman integral is
\be
F_{\Gamma} (q^2,m^2;\ep) = \int \frac{\dd^d k}{(k^{2} -m^{2})^2
(q-k)^{2}} \;.
\ee
The causal $i 0$ in the propagators $k^2-m^2+ i 0$, etc. are omitted
for brevity.

According to (\ref{AE}) two subgraphs give non-zero contributions.
The graph $\Gm$ itself generates Taylor expansion of the integrand in $m$,
with resulting massless integrals evaluated (e.g. by Feynman parameters)
in gamma functions for general $\ep$:
\bea
\int\frac{\dd^d k}{(q-k)^{2}}
 {\cal T}_{m} \frac{1}{(k^{2} -m^{2})^2}
%\nn \\ && \hspace*{-52mm}
&=& \int \frac{\dd^d k}{(k^{2})^2 (q-k)^{2}} - 2 m^2
\int \frac{\dd^d k}{(k^{2})^3 (q-k)^{2}} + \ldots
 \nn \\
%\hspace*{-100mm}
&=& \frac{\I \pi^{d/2}}{(-q^2)^{1+\ep}} \frac{\Gm(1-\ep)^2\Gm(\ep)}{\Gm(1-2\ep)}
\left[1+2\ep\frac{m^2}{q^2}+\ldots \right] \; .
\eea

The second contribution originates from the subgraph $\gm_1$
which is the upper line. It is given by
Taylor expansion of its propagator in the loop momentum $k$
which is external for this subgraph,
with resulting massive vacuum integrals also evaluated
in gamma functions for general $\ep$:
\bea
\int\frac{\dd^d k}{(k^{2} -m^{2})^2}
 {\cal T}_{k} \frac{1}{(q-k)^{2}}
%\nn \\ && \hspace*{-52mm}
&=& \frac{1}{q^2} \int \frac{\dd^d k}{(k^{2} -m^{2})^2}
+ \frac{1}{(q^2)^2} \int \frac{(2q\cdo k-k^2) \dd^d k}{(k^{2} -m^{2})^2}
+ \ldots
\nn \\
 %\hspace*{-100mm}
&=&\frac{\I \pi^{d/2}}{q^2(m^2)^{\ep}}
\Gm(\ep)\left[1+\frac{\ep}{1+\ep} \frac{m^2}{q^2} + \ldots \right]\; .
\eea

The contribution of another subgraph consisting of two lower lines
generates a zero contribution because this is a massless vacuum
diagram:
\[ \int\frac{\dd^d k}{k^{2}}
 {\cal T}_{k,m} \frac{1}{((q-k)^{2} -m^{2})^2} =
 \frac{1}{(q^2)^2}  \int\frac{\dd^d k}{k^{2}} + \ldots=0 \;.
\]
This contribution would be however non-zero in the case of
a non-zero mass in the lower lines.

When $\ep\to 0$, infrared (IR) poles in the first non-zero contribution are canceled
against ultraviolet (UV) poles in the second one, with the finite result
\be
F_{\Gamma} (q^2,m^2;0)  \sim
\frac{\I \pi^2}{q^2} \left[ \ln \left(\frac{-q^2}{m^2}\right)
-\frac{m^2}{q^2}+\ldots \right] \; .
\ee

It turns out that at present there are no simple generalizations of
the strategy of subgraphs to typical Minkowskian regimes.\footnote{
With the exception of
the large momentum off-shell limit and one of the versions of the
Sudakov limit  --- see \cite{acvs}.}
Before formulating what the strategy of regions is let us remind that
a ({\em standard}) strategy of regions was used for many years
for analyzing leading power and (sub)leading logarithms. It reduces to the
following prescriptions:
\begin{itemize}
\item
Consider various regions of the loop momenta
%that are typical for the given limit
and expand, in every region,
the integrand in a Taylor series with respect to the parameters
that are considered small in the given region;
\item
pick up the leading asymptotic behaviour generated by every
region.
\end{itemize}
Let us stress that cut-offs that specify the regions are not removed
within this strategy. In fact, it was sufficient to analyze rather
limited family of regions because the leading
asymptotics are generated only by specific regions.

The ({\em generalized}) strategy of regions
has been suggested in \cite{BS} (and immediately applied to
the threshold expansion):
\begin{itemize}
\item
Consider various regions  \ldots %of the loop momenta
%that are typical for the given limit
%and expand, in every region,
%the integrand in a Taylor series with respect to the parameters
%that are considered small in the given region;
\item
Integrate the integrand expanded, in every region in its own way,
over the whole integration domain in the loop momenta;
\item
Put to zero any integral without scale.
\end{itemize}

Let us stress that, for typically Euclidean limits, integrals without
scale (tadpoles) are {\em automatically} put to zero.
For general limits, this is an {\em ad hoc} prescription.

An experimental observation tell us that
this strategy of regions gives asymptotic expansions
for any diagram in any limit.
In particular, it has been checked in numerous examples when comparing
results of expansion with existing explicit analytical results.
We have also an indirect confirmation because,
for limit typical for Euclidean space, the strategy of regions leads to the same
prescriptions as the strategy of subgraphs. To see this it is in fact
sufficient to take any loop momentum to be either
\bea
large: && k\sim q\;,\;\;\;\;\mbox{or} \nn \\
small: && k\sim m  \nn
\eea
and then observe that one obtains eq.~(\ref{AE}).

Still to see how the strategy of regions works let us consider the
previous example of Fig.~\ref{fig1}. We consider the loop momentum
$k$ to be either large or small and obtain
\bea k \; \mbox{large:}& \ria& {\cal T}_{m}
\frac{1}{(k^2-m^2)^{2}} \lra \Gm
 \nn \\
k \; \mbox{small:} &\ria & {\cal T}_{k} \frac{1}{(q-k)^{2}} \lra \gm_1
\nn
\eea
Thus the region of the large momenta reproduces the contribution
of the subgraph $\Gm$ and the region of the small momenta reproduces
the contribution of the subgraph $\gm_1$ present according to
the strategy of subgraphs (\ref{AE}).

From now on we turn to various examples of limits typical for
Minkowski space.

\section{Strategy of regions for limits
typical for Minkowski space}
%\subsection{Typical for pseudo-Euclidean space}
%

\subsection{Threshold expansion \cite{BS}}

Consider first the threshold limit when an external momentum squared
tends to a threshold value. Our primary task is to see what
kinds of regions are relevant here. Let us consider the same example
of Fig.~\ref{fig1} but in the new limit, $q^2 \to m^2$.
In this case, it is reasonable to choose the loop momentum in another
way to make explicit the dependence on the expansion parameter:
\be
F_{\Gamma} (q^2,y;\ep) = \int \frac{\dd^d k}{k^{2}
((q-k)^{2} -m^{2})^2} %\nn \\
= \int \frac{\dd^d k}{k^{2}
(k^2 -2q\cdo k -y)^2} \;.
\ee
So we have turned to the new variables
$(q^2,m^2) \ria (q^2,y)$ with $y=m^2-q^2\to 0$ the expansion parameter
of the problem.

Let us look for relevant regions.
The region of large (let us
from now on use the term {\em hard} instead)
 momenta, $k\sim q$, always contributes. It gives
\be
\int \frac{\dd^d k}{k^{2}}
{\cal T}_y\frac{1}{(k^2 -2q\cdo k -y)^2} =
\int \frac{\dd^d k}{k^{2}(k^2 -2q\cdo k )^2} +\ldots
= \frac{\I \pi^{d/2}}{(q^2)^{1+\ep}}
\frac{\Gm(1+\ep)}{2\ep}+\ldots ,
\ee
where each integral is evaluated in gamma functions for general $\ep$.

If we consider the region of small loop momenta, $k\sim \sqrt{y}$
(which from now on we will call {\em soft}) we shall obtain
an integral without scale  which we put to zero according to
one of the prescriptions of the strategy of regions:
\[ \int \frac{\dd^d k}{k^{2}(-2q\cdo k)^2} +\ldots =0\;. \]

It is the {\em ultrasoft} (us)
region, $k\sim y/\sqrt{q^2}$, which gives here the second
non-zero contribution:
\be
\int \frac{\dd^d k}{k^{2}}
{\cal T}_{k^2} \frac{1}{(k^2 -2q\cdo k -y)^2} =
\int \frac{\dd^d k}{k^{2}(-2q\cdo k -y)^2} +\ldots \nn \\
= -i \pi^{d/2}  \frac{\Gm(1-\ep)\Gm(2\ep)}{(q^2)^{1-\ep}y^{2\ep}}\;.
\ee
Only the leading term survives because, in the next terms
the factor $k^2$ resulting from expansion cancels the massless
propagator so that a scaleless integral appears.

If we combine the hard and ultrasoft contributions we shall obtain,
in the limit $\ep\to 0$,
the known explicit result for the given diagram expanded
at threshold:
\[
\frac{i \pi^{d/2}}{q^2} \left[ \ln \frac{y}{q^2}
- \frac{y}{q^2} +\ldots \right]\;.
\]

\begin{figure}[b!]
\begin{center}
\epsfig{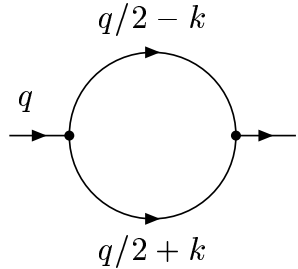}
\caption[0]{\label{f2} One-loop propagator diagram with two non-zero
masses in the threshold}
\label{fig2}
\end{center}
\end{figure}

It turns out that for diagrams consisting of massless and massive
(with the same mass $m$) lines and having thresholds only with one
massive line, i.e. at $q^2=m^2$, only hard and ultrasoft regions
are relevant. To find other characteristic regions we turn
to an example with two massive lines --- see Fig.~\ref{fig2}.
%%%%%%%%%%%%%%%%%%%%%%%%%%%%%%%%%%%%%%%%%%%%%%%%%%%%%%%%%%%%%%
We have
\be
F_{\Gamma} (q^2,y;\ep) = \int \frac{\dd^d k}{(k^{2}-m^2)
((q-k)^{2} -m^{2})} %\nn \\
= \int \frac{\dd^d k}{(k^2 +q\cdo k -y)
(k^2 -q\cdo k -y)}\;,
\ee
where the loop momentum in again chosen in another way, and we have
turned to the new variables:
$(q^2,m^2) \ria (q^2,y)$ where $y=m^2-q^2/4\to 0$ is the
small parameter of the problem.
Keeping in mind the non-relativistic flavour of the problem
we choose the frame $q=\{q_0,\vec{0}\}$.

Let us look for relevant regions. The hard region, $k\sim q$, gives
\bea
\int \dd^d k
{\cal T}_y\frac{1}{(k^2 +q_0 k_0 -y)
(k^2 -q_0 k_0 -y)}
 +\ldots   %&&\nn \\ && \hspace*{-70mm}
&=&\int \dd^d k
\frac{1}{(k^2 +q_0 k_0 )
(k^2 -q_0 k_0)}+\ldots \nn \\
&=& \I \pi^{d/2} \left(\frac{4}{q^2}\right)^{\ep}
\frac{\Gm(\ep)}{1-2\ep} +\ldots
\eea

The soft and ultrasoft regions generate zero contributions
because of the appearance of scaleless integrals:
\be
- \frac{1}{q^2}\int \frac{\dd^d k}{k_0^2} +\ldots =0 \;, %\\
\;\;\; -\frac{1}{q^2}\int
\frac{\dd k_0 \dd^d \vec{k}}{(q_0 k_0 -y+i 0)(q_0 k_0 +y-i 0)}
+\ldots =0\;.
\ee

It turns out that the missing non-zero contribution here comes
from the {\em potential} (p) \cite{BS} region,  $k_0\sim y/q_0\, , \;
\vec{k}\sim \sqrt{y}$. It generates Taylor expansion in $k_0^2$
and is evaluated by closing the integration contour
in $k_0$ and taking a residue, e.g. in the upper half-plane, and
then evaluating $(d-1)$-dimensional integral in $\vec{k}$ using
Feynman parameters.
Here again only the leading term survives because
the next terms involve scaleless integrals:
\bea
\int\dd k_0 \dd^{d-1} \vec{k}{\cal T}_{k_0^2}
\frac{1}{(k^2+q_0 k_0 -y+i 0)
(k^2-q_0 k_0 -y+i 0)} && \nn \\ && \hspace*{-115mm}
=\int\frac{\dd k_0 \dd^{d-1} \vec{k}}{(\vec{k}^2-q_0  k_0 +y-i 0)
(\vec{k}^2+q_0 k_0 +y-i 0)}
+\ldots
%\nn\\ &=&
= \I \pi^{d/2}\Gm(\ep-1/2)
\sqrt{\frac{\pi y}{q^2}}y^{-\ep} \;.
\eea
The sum of the hard and potential contributions successfully reproduces
the known analytical result for the given diagram.

\begin{figure}[b!]
\begin{center}
\epsfig{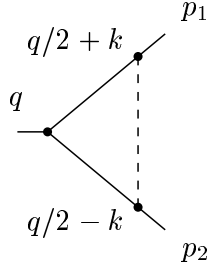}
\caption[0]{\label{f3} Triangle diagram with two non-zero
masses in the threshold}
\label{fig3}
\end{center}
\end{figure}

The next example is given by the triangle diagram with two non-zero
masses in the threshold --- see Fig.~\ref{fig3}.
%%%%%%%%%%%%%%%%%%%%%%%%%%%%%%%%%%%%%%%%%%%%%%%%%%%%%%%%%%%%%%
It is considered at $q=p_1+p_2,\; p=(p_1-p_2)/2,\; p_1^2=p_2^2=m^2$
and is given by the following Feynman integral:
\be \int \frac{\dd^d k}{(k^2 +q\cdo k -y)
(k^2 -q\cdo k -y)(k-p)^2} \;,
\ee
where again $y=m^2-q^2/4\to 0$ and $q=\{q_0,\vec{0}\}$.

The situation is quite similar to the previous diagram. There are two
non-zero contributions generated by the hard and potential regions
\cite{BS}:
the (h) contribution
\be
\int \dd^d k
\frac{1}{(k^2 +q_0 k_0 )
(k^2 -q_0 k_0)(k-p)^2} %\nn \\ && \hspace*{-70mm}
=- \I \pi^{d/2} \left(\frac{4}{q^2}\right)^{1+\ep}
\frac{\Gm(\ep)}{2(1+2\ep)}+\ldots
\ee
and the (p) contribution
\be
\int\frac{\dd k_0 \dd^{d-1} \vec{k}}{(\vec{k}^2-q_0 k_0 +y-i 0)
(\vec{k}^2+q_0 k_0 +y-i 0)(-(\vec{k}-\vec{p})^2)}
%&&\nn\\ && \hspace*{-70mm}
= \I \pi^{d/2}  \frac{y^{-\ep}}{\sqrt{q^2 y}}
\frac{\sqrt{\pi}\Gm(\ep+1/2)}{2\ep}\;.
\ee
One can check that their sum equals the whole analytical result
for the given diagram.

\begin{figure}[b!]
\begin{center}
\epsfig{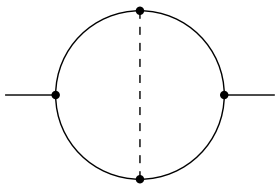}
\caption[0]{\label{f4} Two-loop master self-energy diagram with
two non-zero masses in the threshold}
\label{fig4}
\end{center}
\end{figure}

It turns out that we have already seen the whole list of regions
relevant to the threshold expansion, with the qualification that
soft regions did not yet contribute in the examples.
We refer for two-loop examples to \cite{BS}. For example,
the threshold expansion of Fig.~\ref{fig4} at $y=m^2-q^2/4\to 0$
consists of
%%%%%%%%%%%%%%%%%%%%%%%%%%%%%%%%%%%%%%%%%%%%%%%%%%%%%%%%%%%%%%
contributions generated by the following regions:
(h-h), (h-p)=(p-h), (p-p), (p-us) (where two loop momenta are
characterized, and the ultrasoft momentum in the last contribution
refers to the momentum of the middle line).

\begin{figure}[b!]
\begin{center}
\epsfig{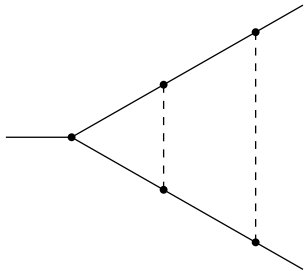}
\caption[0]{\label{f5} Two-loop vertex diagram with
two non-zero masses in the threshold}
\label{fig5}
\end{center}
\end{figure}

Similarly,
the threshold expansion of Fig.~\ref{fig5} at $y=m^2-q^2/4\to 0$
%%%%%%%%%%%%%%%%%%%%%%%%%%%%%%%%%
consists of
(h-h), (h-p), (p-h), (p-p), (p-s)
(where the loop momentum of the box subgraph is soft)
and (p-us) (where the momentum of the middle line is ultrasoft)
contributions --- see
details in \cite{BS}.

It should be stressed that the knowledge about expansions
of individual Feynman diagram gives the possibility to
derive expansions at the operator level.
The threshold expansion with one zero (small) and one non-zero mass
in the threshold leads to HQET
 (see \cite{HQETrev} for review), while
the situation with two non-zero masses in the threshold provides
the transition from QCD to NRQCD \cite{NRQCD}
and then further to pNRQCD  \cite{NRQCD}.
Historically, the development of HQET was performed without the knowledge
of the corresponding diagrammatical expansion. This can be explained
by the combinatorial simplicity of HEQT with the structure similar to that
of the large mass expansion, where there are only two scales in the
problem.

The case of the threshold expansion with two non-zero masses
in the threshold is much more complicated. At the diagrammatical level,
this is described by the multiplicity of relevant regions in the problem
which correspond to three different scales: $m$ (mass of the quark),
the momentum $m v$, where $v$ is relative velocity of the quarks (straightforwardly
expressed through the variable $y$ in the above examples), and
the energy $m v^2$. An adequate description of the transition from NRQCD
(which is obtained from QCD by `integrating out' the hard scale, $m$)
to pNRQCD has been obtained not so easily (see a discussion from the
point of view of 1997 in \cite{BS}), and the development of the
diagrammatical threshold expansion helped to unambiguously identify
all relevant scales in the problem and the form of the corresponding
terms in the effective Lagrangian.

The threshold expansion resulted in a number of applications. The first
of them was analytical evaluation of the two-loop matching coefficients
of the vector current in NRQCD and QCD \cite{CMBSS}.
Another class of important results was the two-loop description
of the  $t\bar{t}$ production in
$e^+ e^-$ annihilation near threshold --- see \cite{Hoang}.

\subsection{Sudakov limit}

\begin{figure}[b!]
\begin{center}
\epsfig{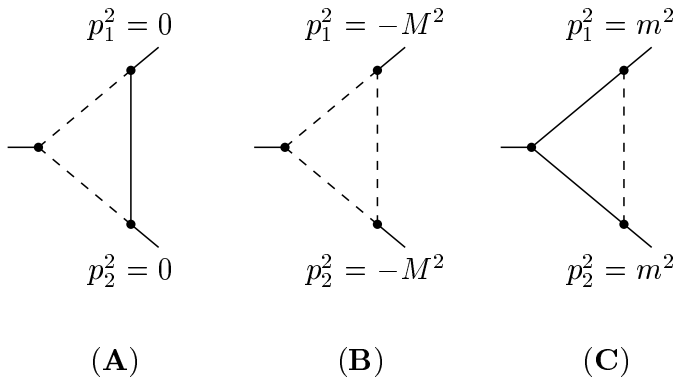}
\caption[0]{\label{f6} Triangle diagram in the Sudakov limit}
\label{fig6}
\end{center}
\end{figure}

There are three different versions of the Sudakov limit
$m^2\ll Q^2\equiv -s=-(p_1-p_2)^2$ or $M^2\ll -s$ which are exemplified
by scalar triangle diagram in Fig.~\ref{fig6}, where dashed lines denote
massless propagators.
%%%%%%%%%%%%%%%%%%%%%%%%%%%%%%%%%%%
Within the `standard' strategy of regions,
summing up (sub)leading logarithms (at the leading power) using
evolution equations has been analyzed in a large number of papers
\cite{SudST}.

Let us expand the triangle diagram
in Limit~A by the strategy of regions. With the standard choice $p_{1,2} = (Q/2,0,0,\mp Q/2)$, we
have
\bea
\int \frac{\dd^dk}{(k^2-2 p_1\cdo k)  (k^2-2 p_2\cdo k) (k^2-m^2)}
&&
\nn \\ && \hspace*{-66mm}
=
\int \frac{ \dd k_+ \dd k_- \dd^{d-2} \uk}
{(k_+k_--\uk^2-Q k_+) (k_+k_--\uk^2-Q k_-)(k_+k_--\uk^2-m^2) } \;,
\nn
\eea
where $k_{\pm} =k_0\pm k_3, \; \uk=(k_1,k_2)$,
with $2 p_{1,2}\cdo k = Q k_{\pm}$.

Let us look for relevant regions. The hard region,
$k\sim q$, generates Taylor expansion of the integrand in $m^2$
\be
\int \frac{\dd^dk}{(k^2-2 p_1\cdo k)(k^2-2 p_2\cdo k) k^2}
+\ldots
%\nn && \\ && \hspace*{-60mm}
= -\I \pi^{d/2} \frac{1}{(Q^2)^{1+\ep}}
\frac{\Gm(1+\ep)\Gm(-\ep)^2}{\Gm(1-2\eps)}
+ \ldots\;.
\ee
The soft  ($k\sim m$) and ultrasoft ( $k\sim m^2/Q$) regions generate
scaleless integrals which are zero:
\be
\int \frac{\dd^dk}{(-2 p_1\cdo k)(-2 p_2\cdo k) k^2}
+\ldots =0 \; , %\\
\;\;\;
\int \frac{\dd^dk}{(-2 p_1\cdo k)(-2 p_2\cdo k) (-m^2)}
+\ldots =0 \;.
\ee

What is yet missing is the contribution of collinear
regions\footnote{introduced within the
`standard strategy of regions' \cite{colreg}}:
\bea
\mbox{{\em 1-collinear} (1c):}
&& k_+ \sim m^2/Q ,\,\,k_-\sim Q\, , \,\, \uk \sim m\,,
\nn
 \\
\mbox{{\em 2-collinear} (2c):} && k_+\sim Q,\,\,k_-\sim m^2/Q\, ,
\,\,\uk \sim m \, .
\nn
\eea
The (1c) region generates Taylor expansion of propagator~2 in $k^2$:
\be
\int \frac{\dd^dk}{(k^2-2 p_1\cdo k)(-2 p_2\cdo k) (k^2-m^2)}
+\ldots,
\ee
and the (2c) contribution is symmetrical. These contributions are
not however individually regularized by dimensional regularization.
A natural way to overcome this obstacle is to
introduce an auxiliary analytic regularization \cite{Sp1},
 calculate (1c) and (2c) contributions and switch it off in the sum.
Then the (1c) and (2c) regions give, in the leading power,
\be
 -\I \pi^{d/2}
\frac{\Gm(\ep) }{Q^2 (m^2)^{\ep}}
%&& \nn \\ && \hspace*{-35mm} \times
\left[ \ln (Q^2/m^2) + \psi(\ep) -\gm_{\rm E}
- 2\psi(1-\ep)  \right] +\ldots
\ee

After we combine the (h) and (c) contributions we shall see that
IR/collinear poles in the (h) contribution and
UV/collinear poles in (c) contribution are canceled, and
 at $\ep\to 0$ we obtain
\be
  \frac{\I\pi^2}{Q^2} \left[
\mbox{Li}_2 ( x ) - \frac{1}{2} \ln^2 x
+ \ln x \ln(1-x) -\frac{ \pi^2}{3} \right] \, ,
\ee
where $\mbox{Li}_2 ( x )$ is dilogarithm and
$x=m^2/Q^2$.

The triangle diagrams of Fig.~\ref{fig5} in
Limits~B and~C are similarly expanded. In Limit~B one meets
(h), (1c), (2c) and (us) regions, and, in Limit~C, one has
(h), (1c) and (2c) regions.

Two-loop examples for the Sudakov limit, within the strategy of regions,
can be found in \cite{SR}.
\begin{figure}[t!]
\begin{center}
\epsfig{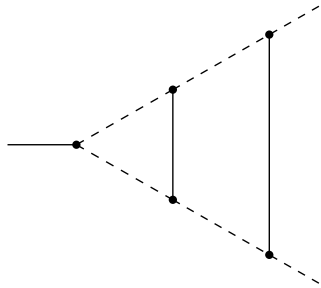}
\caption[0]{\label{fi7} Two-loop vertex diagram in Limit~A}
\label{fig7}
\end{center}
\end{figure}
%%%%%%%%%%%%%%%%%%%%%%%%%%%%%%%%%%%
For example, the following regions contribute to the expansion
of Fig.~\ref{fig7} in Limit~A: (h-h), (1c-h)$+$(2c-h),
(1c-1c)$+$(2c-2c), and (h-s) where the soft momentum refers to the
middle line.
For Limit~B, one has
(h-h), (1c-h)$=$(2c-h), (1c-1c)$=$(2c-2c),
 (us-h), (us-1c), (us-2c), (us-us).

For the diagram of Fig.~\ref{fig5} (considered above in
the threshold limit),
%%
%\begin{figure}[t!]
%\begin{center}
%\epsfig{file=v2sudC.eps,height=2in}
%\epsfig{file=f87C.eps,height=2in}
%\caption[0]{\label{f8} Two-loop vertex diagram in Limit~C.}
%\label{fig8}
%\end{center}
%\end{figure}
%%%%%%%%%%%%%%%%%%%%%%%%%%%%%%%%%%%
with
$p_{1,2} = \tilde{p}_{1,2} + (m^2/Q^2) \tilde{p}_{2,1}\,,$
with $ \tilde{p}_{1,2} = (Q/2,0,0,\mp Q/2)$,
obvious regions
(h-h), (1c-h)$=$(2c-h), (1c-1c)$=$(2c-2c) are not sufficient
because the poles of the fourth order do not cancel. It turns out that
it is necessary to consider also ultracollinear regions:
\bea
\mbox{(1uc):}
&& k_+ \sim m^4/Q^3 ,\,\,k_-\sim m^2/Q\, , \,\, \uk \sim m^3/Q^2\,,
\nn
 \\
\mbox{(2uc):} && k_+\sim m^2/Q,\,\,
k_-\sim m^4/Q^3\, ,
\,\,\uk \sim m^3/Q^2 \, .
\nn
\eea
After one adds contributions of the (1uc-2c) and (1c-2uc) regions
the leading power of expansion satisfies the check of poles \cite{SR}.

The (generalized) strategy of regions combined with
evolution equations derived within the `standard' strategy of regions
has been applied to summing up next-to-leading logarithms for
Abelian  form factor and four-fermion
amplitude in the $SU(N)$ gauge theory \cite{KPS}.

\subsection{Regge limit}

The Regge limit for scattering diagrams is characterized as
$|t| \ll |s|$, where
$s=(p_1+p_2)^2$ and $t=(p_1+p_3)^2$ are Mandelstam variables.
\begin{figure}[b!]
\begin{center}
\epsfig{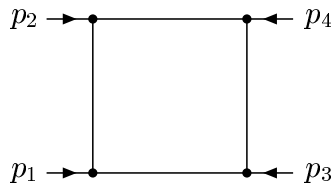}
\caption[0]{\label{f9} Box diagram}
\label{fig9}
\end{center}
\end{figure}
%%%%%%%%%%%%%%%%%%%%%%%%%%%%%%%%%%%

Let us expand, using the strategy of regions, the box diagram
shown in Fig.~\ref{fig9}. The Feynman integral is
\be
\int\frac{\dd^dk}{(k^2+2 p_1 \cdo k) (k^2-2 p_2\cdo  k)
k^2 (k+p_1+p_3)^2} \;.
\ee
Let $s=-Q^2$, $t=-T$, and let us choose
$p_{1,2} = (\mp Q/2,0,0,Q/2)$, and
$p_1+p_3 = (T/Q,0,\sqrt{T+T^2/Q^2},0)$.
It turns out that in the Regge limit one meets contributions
of (h) and (c) regions.
The collinear regions are now characterized as
\[
\mbox{(1(2)c):} \;\;\;
 k_{\pm} \sim T/Q ,\,\,k_{\mp}\sim Q\, , \,\, \uk \sim \sqrt{T} \;.
\]
The sum of (1c) and (2c) gives, in the leading power, $1/t$,
\be
\I\pi^{d/2}
\frac{\Gm(-\ep)^2 \Gm(1+\ep)}{\Gm(-2\ep)s (-t)^{1+\ep}}
% && \nn \\ && \hspace*{-35mm} \times
\left[
\ln (t/s) +\psi(-\ep) -2\psi(1+\ep) +\gm_{\rm E}\right] \;.
%\nn
\ee
The hard contribution starts from the NLO.
If we sum up the (h) and (c) contributions we shall see that,
at $\ep\to 0$,
only the LO (c) contribution survives  and gives
\be
\frac{\I\pi^{d/2}\E^{-\gm_{\rm E} \ep}}{s t}
\left[
\frac{4}{\ep^2} - \left(\ln(-s)+\ln(-t) \right)\frac{2}{\ep}
%\right. && \nn \\  && \hspace*{-35mm} \left.
+2 \ln(-s) \ln(-t) -\frac{4\pi^2}{3}
\right] \;.
%\nn
\ee

In the case  of on-shell massless double box, $p_i^2=0$,
\begin{figure}[t!]
\begin{center}
\epsfig{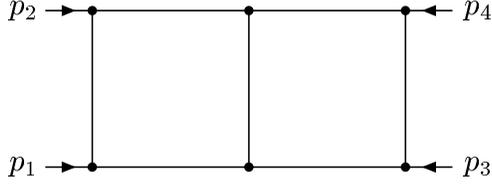}
\caption[0]{\label{f10} Double box}
\label{fig10}
\end{center}
\end{figure}
%%%%%%%%%%%%%%%%%%%%%%%%%%%%%%%%%%%
given by the integral
\bea
\int\int \frac{\dd^dk \dd^dl}{
(l^2+2 p_1 \cdo l)(l^2-2 p_2 \cdo l) (k^2+2 p_1\cdo  k)
(k^2-2 p_2 \cdo k) } &&
\nn \\ && \hspace*{-73mm} \times \frac{1}{k^2 (k-l)^2
(l+r)^2 }
%\nn \\ && \hspace*{-75mm}
\equiv
\frac{\left(\I\pi^{d/2} \E^{-\gm_{\rm E}\ep} \right)^2 }{(-s)^{2+2\ep}(-t)}
\; K(t/s;\ep)\;,
\eea
there are (h-h), (1c-1c) and (2c-2c) contributions in the Regge limit.
The (h-h) contribution starts from the NLO, $t^0$, and
the (c-c) contribution from LO, $t^{-1}$.
This is a result for the sum of the LO and NLO contributions
\cite{SV}
\bea
K(x,\ep) &=&
-\frac{4}{\ep^4} +\frac{5\ln x}{\ep^3}
- \left( 2 \ln^2 x -\frac{5}{2} \pi^2  \right) \frac{1}{\ep^2}
\nn \\ &&  \hspace*{-20mm}
-\left( \frac{2}{3}\ln^3 x +\frac{11}{2}\pi^2 \ln x
-\frac{65}{3} \zeta(3) \right) \frac{1}{\ep}
%\nn \\ && \hspace*{-20mm}
+\frac{4}{3}\ln^4 x +6 \pi^2 \ln^2 x
-\frac{88}{3} \zeta(3)\ln x +\frac{29}{30}\pi^4
\nn \\ && \hspace*{-20mm}
+ 2 x \left[
\frac{1}{\ep}\left(\ln^2 x - 2 \ln x +\pi^2+2\right)
-\frac{1}{3} \left(
4\ln^3 x +3 \ln^2 x
\right.
\right.\nn \\ && \hspace*{-20mm}
\left.\left.
+ (5\pi^2-36)\ln x +2 (33+5\pi^2-3\zeta(3))
\right)
\right]  + O(x^2 \ln^3 x) \;.
\eea

The on-shell double box has provided a curious example of
a situation when the evaluation of large number terms of the
expansion is rather complicated while an explicit analytical
result\footnote{See \cite{Ge} for a review of recent results on the
evaluation of double box diagrams.}
 \cite{SmDB} is known:
\bea
K(x,\ep) &= &
-\frac{4}{\ep^4} +\frac{5\ln x}{\ep^3}
- \left( 2 \ln^2 x -\frac{5}{2} \pi^2  \right) \frac{1}{\ep^2}
\nn \\ && \hspace*{-16mm}
-\left( \frac{2}{3}\ln^3 x +\frac{11}{2}\pi^2 \ln x
-\frac{65}{3} \zeta(3) \right) \frac{1}{\ep}
%\nn \\ && \hspace*{-16mm}
+\frac{4}{3}\ln^4 x +6 \pi^2 \ln^2 x
-\frac{88}{3} \zeta(3)\ln x +\frac{29}{30}\pi^4
\nn \\ && \hspace*{-16mm}
- \left[
2 \Li{3}{ -x } -2\ln x \Li{2}{ -x }
-\left( \ln^2 x +\pi^2 \right) \ln(1+x)
\right] \frac{2}{\ep}
\nn \\ && \hspace*{-16mm}
- 4 \left(S_{2,2}(-x) - \ln x S_{1,2}(-x)  \right)
+ 44 \Li{4}{ -x }
%\nn \\ && \hspace*{-16mm}
- 4 \left(\ln(1+x) + 6 \ln x  \right) \Li{3}{ -x }
\nn \\ && \hspace*{-16mm}
+ 2\left( \ln^2 x +2 \ln x \ln(1+x) +\frac{10}{3}\pi^2\right) \Li{2}{-x}
\nn \\ && \hspace*{-16mm}
+\left( \ln^2 x +\pi^2 \right) \ln^2(1+x)
%\nn \\ && \hspace*{-16mm}
-\frac{2}{3} \left(4\ln^3 x +5\pi^2 \ln x -6\zeta(3)\right) \ln(1+x) \;.
\label{2box}
\eea
 Still the evaluation of those first
two terms of the expansion was used as a very crucial check of
(\ref{2box}). Observe that the asymptotic expansion within the strategy
of regions was successfully applied in \cite{SmDBp} also to double boxes
with one leg off shell.

\section{Present status of the strategy of regions}

To characterize the present status of the strategy of regions let
us first point out  that at present there are
no mathematical proofs, similar to the case of the strategy of subgraphs
(applied only to the limits typical for Euclidean space), although this
looks to be a very good mathematical problem.
(Moreover,
the very word `region' is understood in the physical sense so that
one does not bother about `the decomposition of unity'.)
Its solution is expected to be specific for each concrete regime typical
for Minkowski space. Another reasonable problem is to develop the strategy
of regions for phase space integrals arising in evaluation of
real radiation processes.

Let us conclude with advice that could be useful
when studying a new limit:
\begin{itemize}
\item
Look for regions, typical for the limit (probably, they are
similar to regions connected with known limits\footnote{As a recent example of using the strategy
of regions in a new situation let us refer to ref.~\cite{CM2} where
non-relativistic integrals describing bound states within NRQCD
were
further expanded in the ratio of the small and the large mass $m/M$.
The relevant regions turned out to be (non-relativistic) hard and soft
regions of three-dimensional momenta.});
\item
Test one- and two-loop examples by comparison with
explicit results;
\item Check poles in $\ep$; if this check is not satisfied look for
missing regions;
\item Check expansion numerically;
\item Use the strategy of regions formulated in $\al$-parameters
\cite{SR}, e.g., to avoid double counting;
\item
Stay optimistic because, up to now,
the strategy of regions successfully worked in all known examples!
\end{itemize}

%%%%%%%%%%%%%%%%%%%%%%%%%%%%%%%%%%%%%%%%%%%%%%%%%%%%%%%%%%%%%%%%%%%%%%%%%
%%
%%   use this format to include an .eps figure into your paper
%%
%\begin{figure}[b!]
%\begin{center}
%\epsfig{file=hmhtbmt.ps,height=3in}
%\epsfig{file=f11.eps,height=1in}
%\caption[0]{\label{mhtanb} Caption.}
%\label{fig:higgsmass}
%\end{center}
%\end{figure}
%%%%%%%%%%%%%%%%%%%%%%%%%%%%%%%%%%%%%%%%%%%%%%%%%%%%%%%%%%%%%%%%%%%%%%%%%%%

\Acknowledgments I am thankful to S.~Brodsky, L.~Dixon, H.~Haber
and  K.~Melnikov for support and kind hospitality at the symposium.


\begin{thebibliography}{99}

%%
%%  bibliographic items can be constructed using the LaTeX format in SPIRES:
%%    see    http://www.slac.stanford.edu/spires/hep/latex.html
%%  SPIRES will also supply the CITATION line information; please include it.
%%

\bibitem{BS}
M.~Beneke and V.A.~Smirnov, {\sl Nucl. Phys.} {\bf B522} (1998) 321.

\bibitem{SS}
S.G.~Gorishny, {\sl Nucl. Phys.} {\bf B319} (1989) 633;
K.G.~Chetyrkin, {\sl Teor. Mat. Fiz.} {\bf 75} (1988) 26; {\em ibid.}
{\bf 76} (1988) 207;
V.A.~Smirnov, {\sl Commun. Math. Phys.} {\bf 134} (1990) 109.

\bibitem{Sm3}
V.A.~Smirnov, {\sl Mod. Phys. Lett.} {\bf A10} (1995) 1485.


\bibitem{dimreg}
G.~'t Hooft and M.~Veltman, {\sl Nucl.~Phys.}  {\bf B44} (1972) 189;
C.G.~Bollini and J.J.~Giambiagi, {\sl Nuovo Cim.}  {\bf 12B} (1972) 20.


\bibitem{CKK}
K.G.~Chetyrkin, J.H.~K\"uhn and A.~Kwiatkowski,
 {\sl Phys. Reports} {\bf 277} (1997) 189.

\bibitem{acvs}
V.A.~Smirnov, {\sl Phys. Lett.} {\bf B394} (1997) 205; {\bf B404} (1997)
101;
A.~Czarnecki and V.A.~Smirnov, {\sl Phys. Lett.} {\bf B394} (1997) 211.

\bibitem{HQETrev}
M. Neubert, {\sl Phys. Rep.} {\bf 245} (1994) 259.

\bibitem{NRQCD}
W.E.~Caswell and G.P.~Lepage, {\sl Phys. Lett.} {\bf B167} (1986) 437.

\bibitem{pNRQCD}
A.~Pineda and J.~Soto, {\sl Nucl. Phys. Proc. Suppl.} {\bf 64} (1998)
428; {\sl Phys. Rev.} {\bf D59} (1999) 016005.

\bibitem{CMBSS}
A.~Czarnecki and K.~Melnikov, {\sl Phys. Rev. Lett.} {\bf 80}
(1998) 2531;
M.~Beneke, A.~Signer and V.A.~Smirnov, {\sl Phys. Rev. Lett.} {\bf 80}
(1998) 2535.

\bibitem{Hoang}
A.~Hoang et al., {\sl Eur. Phys. J. direct}  {\bf C3} (2000) 1;
A.~Hoang, these proceeidngs.

\bibitem{SudST}
V.V.~Sudakov,  {\sl Zh. Eksp. Teor. Fiz.} {\bf 30} (1956) 87;
R.~Jackiw, {\sl Ann. Phys.} {\bf 48} (1968) 292; {\bf 51} (1969) 575;
J.M.~Cornwall and  G.~Tiktopoulos, {\sl Phys. Rev. Lett.}
{\bf 35} (1975) 338; {\bf D13} (1976) 3370;
J.~Frenkel and J.C.~Taylor, {\sl Nucl. Phys.}
{\bf B116} (1976) 185;
J.C.~Collins, {\sl Phys.Rev.} {\bf D22} (1980) 1478;
in {\em Perturbative QCD}, ed. A.H.~Mueller, 1989, p.~573.
A.~Sen, {\sl Phys. Rev.} {\bf D24} (1981) 3281; {\bf D28} (1983) 860;
G.~Korchemsky, {\sl Phys. Lett.}  {\bf B217} (1989) 330;
{\bf B220} (1989) 629.

\bibitem{colreg}
G.~Sterman, {\sl Phys. Rev.} {\bf D17} (1978) 2773;
S.~Libby and G.~Sterman, {\sl Phys. Rev.} {\bf D18} (1978) 3252;
A.H.~Mueller, {\sl Phys. Rep.} {\bf 73} (1981) 35.

\bibitem{Sp1}
E.R.~Speer, {\sl J. Math. Phys.} {\bf  9} (1968) 1404.

\bibitem{SR}
V.A.~Smirnov and E.R.~Rakhmetov,  {\sl Teor. Mat. Fiz.} {\bf 120}
(1999) 64;
V.A.~Smirnov, {\sl Phys. Lett.}  {\bf B465} (1999) 226.

\bibitem{KPS}
J.H.~K\"uhn, A.A.~Penin and V.A.~Smirnov, {\sl Eur. Phys. J.}
{\bf C17} (2000) 97.

\bibitem{SV}
V.A.~Smirnov and O.L.~Veretin,
{\sl Nucl. Phys.} {\bf B566} (2000) 469.

\bibitem{SmDB}
V.A.~Smirnov, {\sl Phys. Lett.} {\bf B460} (1999) 397.

\bibitem{Ge}
T.~Gehrmann, these proceedings.

\bibitem{SmDBp}
V.A.~Smirnov, {\sl Phys. Lett.} {\bf B491} (2000) 130;
V.A.~Smirnov, hep-ph/0011056.

\bibitem{CM2}
A.~Czarnecki and K.~Melnikov,  hep-ph/0012053.

\end{thebibliography}
\end{document}